# High-flux water desalination with interfacial salt sieving effect in nanoporous carbon composite membranes

Wei Chen[1], Shuyu Chen[2], Qiang Zhang[1], Zhongli Fan[1], Kuo-Wei Huang[1], Xixiang Zhang[1], Zhiping Lai[1*], Ping Sheng[2,3*]

[1]*Division of Physical Science and Engineering, King Abdullah University of Science and Technology, Thuwal, Saudi Arabia*
[2]*Department of Physics, Hong Kong University of Science and Technology, Clear Water Bay, Kowloon, Hong Kong, China*
[3]*Institute for Advanced Study, Hong Kong University of Science and Technology, Clear Water Bay, Kowloon, Hong Kong, China*

**Nanoporous carbon composite membranes, comprising a layer of porous carbon fiber structures with an average channel width of 30-60 nm grown on a porous ceramic substrate, are found to exhibit robust desalination effect with high freshwater flux. In three different membrane processes of vacuum membrane distillation, reverse osmosis and forward osmosis, the carbon composite membrane showed 100% salt rejection with 3.5 to 20 times higher freshwater flux compared to existing polymeric membranes. Thermal accounting experiments found that at least 80% of the freshwater pass through the carbon composite membrane with no phase change. Molecular dynamics simulations revealed a unique salt rejection mechanism. When seawater is interfaced with either vapor or the surface of carbon, one to three interfacial atomic layers contain no salt ions. Below the liquid entry pressure, the salt solution is stopped at the openings to the porous channels and forms a meniscus, while the surface layer of freshwater can feed the surface diffusion flux that is fast-transported on the surfaces of the carbon fibers, driven by the chemical potential gradient. As the surface-transported water does not involve a phase change, hence that component involves no energy expenditure in the form of latent heat.**

*Emails: zhiping.lai@kaust.edu.sa, sheng@ust.hk

**Introduction**

With the dwindling of surface and underground freshwater resources, water desalination has been thrust into an increasingly important role as a means of supplying freshwater to a thirsty world in which the increasing trend of water usage is not balanced by the available supply[1-3]. Traditional desalination approaches involve either the process of distillation which needs large amount of energy, or the filtration approach using polymeric membranes which need to achieve both high salt rejection rate and high freshwater flux. Most desalination units installed in the last twenty years have adopted the energetically more efficient reverse osmosis (RO) membrane process[3,4], while membrane distillation (MD) and forward osmosis (FO) have also attracted intense attention in recent years[5] because of their potential integration with renewable energies. In all the filtration approaches, membrane flux constitutes a common challenge. High flux membrane is highly desirable not only for reducing the membrane area, but also for increasing the productivity.

Existing polymeric membranes used in the reverse osmosis (RO) and forward osmosis (FO) desalination processes are necessarily dense so as to achieve high salt rejection rate, whereas for the membrane distillation (MD) process the membrane can be microporous for the transport of water vapor. In all the polymeric membrane processes the flux is limited by either the low permeability of the dense membrane or the low density of the transported water vapour. Water was found to transport superfast along aquaporin[6-8] and carbon nanotube[2,9-19] channels. An obvious approach to improving the membrane flux is by embedding either of these two materials into a dense matrix to form matrix membranes[20-36]. Indeed, a commercial aquaporin membrane made by this approach gives a high water flux of 7 liters per square meter per hour (LMH) in the FO processes[37]; while a carbon nanotube/polyamide composite membrane was able to significantly improve the specific water flux up to 3.6 LMH/bar in the RO processes[38,39]. However, the mixed matrix membrane approach suffers from many challenges such as poor dispensability, low loading rate, improper alignment and defects, etc. Moreover, although aligned carbon nanotube layers showed enhanced water flux[40-42], but to the best of our knowledge their effect in water desalination has not yet been clearly demonstrated. In particular, to achieve a good salt rejection rate, the pore size of the carbon nanotubes needs to be less than 1.1 nm, as indicated by a number of simulation studies[20-24]. Very recently, graphene membranes showed promising potentials in gas and liquid separations[43-48], but their application in desalination still remains in the stage of theoretical predictions[45,48].

Here we report the synthesis of a nanoporous carbon composite membrane containing a layer of carbon fibers on porous ceramic support; it has a relatively open structure with a minimum pore size of ~30 nm. The membrane was successfully applied to all three membrane desalination processes and showed 100% salt rejection with 3.5 to 20 times higher freshwater flux when compared to existing polymeric membranes; the only limitation with the present desalination approach is that the applied pressure cannot exceed the liquid entry pressure of our membrane, hence the RO process is applicable only to brackish water desalination. But no such limitation exists for the other two desalination processes. From a combination of VMD, FO, and energy accounting experiments, water was found to transport along the outer wall of the carbon fibers. From molecular dynamics simulations, a novel interfacial salt sieving effect is discovered that can account for the high salt rejection rate. This interfacial salt sieving effect is shown to differ fundamentally from the solution-diffusion mechanism in polymeric membranes; it also differs from the molecular sieving mechanism that is expected in carbon nanotubes and graphitic materials. Instead of requiring very small nanopores to prevent the passing of salt ions, our membrane is relatively open in structure. The desalination mechanism arises from the interfacial freshwater layer whose existence is owed to the well-known fact that the salt ions are always individually enveloped by water molecules (the so-called solvation shells), thereby preventing them from direct contact with the graphitic surfaces. This interfacial freshwater layer can feed the surface diffusion flux that is fast-transported on carbon surfaces. Details of a consistent physical picture are given below. Our membrane performance is robust and insensitive to defects, with greatly reduced difficulty in membrane preparation and improved process reproducibility.

**Membrane structure**

Our carbon composite membrane was fabricated on a hollow yttrium-stabilized zirconia (YSZ) tube (Fig. 1(a)) with a porous wall. The obtained composite membrane is denoted as C-DP-X, where P denotes the nickel deposition power in Watts and X denotes the growth time in minutes.

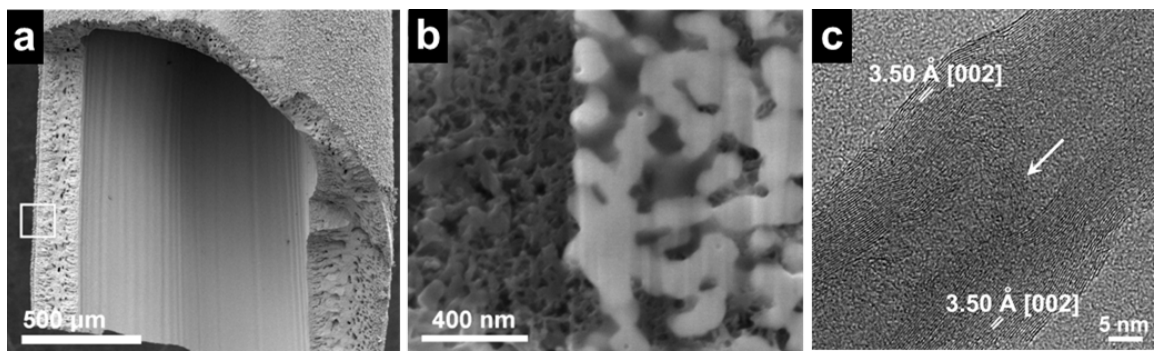

**Figure 1 | Structure of the membrane. (a)** SEM image of an as-prepared C-D35-2 membrane on the surface of the hollow YSZ tube. The square denotes the area to be zoomed in for a magnified view. **(b)** The FIB-SEM image of the interface between YSZ and carbon layer. The sharp interface between carbon and YSZ is clearly delineated. The nano-sized pores on the carbon side can also be seen. The pore size is seen to be the smallest in the vicinity of the carbon fiber-ceramic interface, about 31 nm as determined by gas permeation. **(c)** HRTEM image of a typical single carbon nano-fiber in the C-D35-2 membrane. The arrow points to the "bamboo-knot-like" structure inside the carbon fiber that divides the interior space into compartments.

The typical membrane structure is shown in Fig. 1. The thickness of the entire carbon layer is about 10 μm with a loose outer surface and a dense interface that separates the carbon layer and YSZ support, as evidenced in Fig. 1(b) as well as by EDX mapping analysis (see Fig. S2 in Supplementary Information). The effective region of the membrane, consisting of a dense growth of carbon fibers at the interface with the YSZ substrate, is only about 0.5 μm in thickness. The average pore size, as determined by gas permeation[49], is 31 nm. Detailed studies of many carbon fibers by high resolution transmission electron microscopy, one of which is shown in Fig. 1(c), and Raman spectrum (Fig. S3 in Supplementary Information) revealed that every fiber studied has a multiwall carbon nanotube structure, but the inner channels are always blocked by bamboo-knot-like structures, indicated by an arrow in Fig. 1(c).

**Liquid entry pressure and flux transport via Knudsen and surface diffusion**

Water cannot penetrate through the nanoporous carbon membrane unless an applied pressure is higher than the liquid entry pressure (LEP). The measured LEPs for membranes with different pore sizes follow accurately the relation LEP=4 $\gamma |\cos\theta|$ /(pore size) (see Supplementary Information and Fig. S4), with θ = 93° being obtained by fitting the experimental values of LEP with the simulated water-vapour interfacial tension of 58 mN/m (see below). Therefore, the nanoporous carbon membrane is equivalent to a porous (slightly) hydrophobic membrane that can be used for seawater desalination via membrane distillation at pressures below the LEP. In this regime there are two mechanisms for freshwater transport—Knudsen diffusion of water vapour and surface diffusion of freshwater. In our case the measured flux, at below the LEP, was found to be much higher than that for the Knudsen diffusion, whereas in the comparison cases using the polymeric membranes the measured flux is always in good agreement with the predictions of the Knudsen diffusion. Below we use the MD mode of the desalination process to show that in contrast to the commercial porous polytetrafluoroethylene (PTFE) membrane (W.L. Gore®), our nanoporous carbon membrane can exhibit a very large difference in the freshwater flux between the contact mode of operation in which the water is in contact with the membrane, and the non-contact mode in which only the vapour is in contact. We attribute the very large additional freshwater flux in the contact mode (up to more than 80%) to the surface diffusion mechanism. This is because the strong salt concentration dependence of the feed solution, observed in the contact mode of the MD process (Fig. 2(b)) for our membranes, excludes the Knudsen diffusion as the dominant mechanism, which is nearly independent of the salt concentration in the feed solution. In what follows, the large surface diffusion flux of freshwater in our membrane is shown to be quantitatively compatible with the flux data in the FO process, as well as confirmed by the energy accounting experiments in which the latent heat carried by the vapour offers a way to obtain an upper bound on the transported vapour fraction.

**Membrane distillation**

An illustration of a vacuum membrane distillation (VMD) setup is shown in Fig. 2(a), where a C-D35-2 membrane was immersed into a salt solution with one end sealed by epoxy resin and the other end connected to a vacuum pump through a condensation cold trap. The latter can use either liquid nitrogen or cold water at $2^{\circ}$ C, with the cold water showing only 1% less freshwater collected. NaCl solutions were used as synthetic seawaters. The salt concentration was measured by conductivity at room temperature. In all the experiments the conductivity of the collected water after VMD is less than 2 μS/cm, equivalent to 1 ppm salt concentration. Therefore, the salt rejection rate is over 99.9%. The freshwater fluxes of C-D35-2 membrane, at different temperatures and different salinity of the salt solution, are shown in Fig. 2(b). The flux increases as temperature increases. Above 40 $^{\circ}$C, the water flux increases almost linearly with temperature, while the trend should have followed the exponential relationship if the transport mechanism is dominated by Knudsen diffusion. This is already an indication that the mechanism differs from Knudsen diffusion. At 90 $^{\circ}$C, approximately 1.34 liter freshwater was collected from a 5 wt% NaCl solution after 48 hours over a membrane area of $1.26\times10^{-4}$ m$^2$, which gives a water flux of 221.6 LMH. Reducing the salinity of the feed solution can increase the water flux up to 413.5 LMH when freshwater is used as the feed solution. These values are not only significantly higher than the highest values reported for polymeric membranes, which was around 80 LMH to the best of our knowledge[50], but also 15 to 20 times higher than that obtained by using the non-contact mode, in which only vapour is in contact with the membrane and hence the transport must follow the Knudsen diffusion mechanism. The results are shown by the blue stars in Fig. 2(b). For comparison, we investigated the water flux on a PTFE membrane on the same setup. The measured water fluxes are shown as red open squares in the inset figure. The data match very well with the predictions based on Knudsen diffusion, while the absolute values are 5 to 10 times lower than that of the nanoporous carbon membrane. Based on these observations, we attribute the freshwater flux in our nanoporous carbon membrane that is in excess of the amount obtained by the non-contact mode (i.e., through Knudsen diffusion), to be transported by the surface diffusion mechanism. Such excess amount can account for more than 80% of the total flux. In contrast, all the polymeric membranes used in membrane distillation, such as PTFE, polyvinylidene fluoride (PVDF), polypropylene (PP), etc., display only a negligible amount of surface diffusion since their water flux matches very well with the prediction of the Knudsen equation.

We have also conducted VMD desalination process on seawater taken from the Red Sea, with a salinity of 4.1%. Quantitative details are shown in Figs. S7 and S8 in the Supplementary Information. The nanoporous carbon membrane also showed excellent desalination performance for the real seawater (Fig. S7), rejecting all the monovalent and divalent ions in the seawater.

The freshwater flux is slightly lower when compared to salt solution with the same salinity, owing mainly to the presence of divalent ions in seawater. Hence the freshwater flux for seawater desalination is equivalent to the flux of synthetic seawater of (a slightly higher) 5% salinity. As it is well-known that divalent ions reduce the water flux much more than monovalent ions[51], similar reduction in the freshwater flux has been observed on commercial membranes[52,53]. We have also tested scalability of our desalination approach by using multiple membranes in parallel, and found the total flux to be a linear function of the number of membranes (see Supplementary Information, section (4.2)).

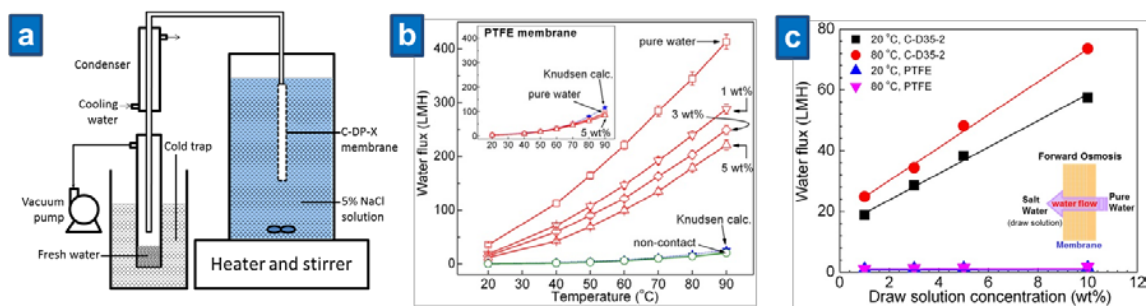

**Figure 2 | Freshwater transport through the C-D35-2 membrane. (a)** A schematic illustration of the VMD setup. **(b)** The measured freshwater flux plotted as a function of temperature (red lines) at different salt concentrations. Blue stars denote the water flux predicted by Knudsen diffusion (labelled as Knudsen calculation), based on the measured data using methane gas, which has very similar molecular weight as water vapour. The green open circles denote the water flux in the non-contact model when the membrane was exposed to only vapour (obtained by bubbling $N_2$ through water); good agreement with the Knudsen diffusion is seen. The inset shows the results over a PTFE membrane (pore size ~100 nm), where the red lines are measured freshwater fluxes at different salt concentrations and the blue line is the calculated flux by Knudsen diffusion. **(c)** The membrane freshwater flux in the FO process at two different temperatures (maintained to be the same on both sides of the membrane) plotted as a function of the draw solution salinity. The freshwater flux for the PTFE membrane is seen to be more than an order of magnitude lower. Inset: A schematic illustration of the FO process.

**Forward and reverse osmosis**

Apart from the VMD mode of desalination, we show in the inset to Fig. 2(c) a FO process in which the membrane separates pure water from the salt solution, denoted the draw solution in general. Driven by the concentration gradient, pure water would diffuse across the membrane to the draw solution. This is evidenced by the linear variation of the pure water flux with the salinity of the draw solution as seen in Fig. 2(c). The magnitude of the water flux for the nanoporous carbon membrane is seen to be more than an order of magnitude higher than that of the PTFE membrane. This huge contrast is significant in demonstrating the advantage of the carbon composite membrane. Moreover, in spite of the salt ion concentration gradient in the reverse direction, the salt leakage rate from the draw solution to the pure water stream was almost zero, as the salt concentration detected in the pure water stream was below 1 ppm during the 2-days FO measurement.

Similar high freshwater flux was measured in the RO process (see Supplementary Information and Fig. S9). A pressure of 3 bars, necessarily less than the LEP of the membrane, was applied to the salt solution side. The salt solution is at the concentration of 2000 ppm, which is in the brackish water salinity range. The applied pressure of 3 bars allowed the extraction of freshwater from salt solution. At $80^\circ$C, the specific water flux of the RO process reached almost 29 LMH/bar with almost 100% salt rejection rate. This is 3.5 times the performance of the best commercial RO polymeric membranes, about 8 LMH/bar under similar salinity and laboratory conditions[3]. Further reduction in the pore size and therefore increasing the LEP of the membrane will allow our desalination approach to use the RO process for higher salinity water.

**Surface diffusion constant**

The substantial amount of surface diffusion indicates that we can accurately determine the surface diffusion constant by using the FO process. The water flux can be well fitted to the following surface diffusion equation:

$$J_w = \varepsilon D \frac{\Delta C_w}{\delta}, \qquad (1)$$

where $J_w$ is the water flux calculated on the basis of total surface area, $\delta$ (=0.5 µm) is the thickness of the dense carbon layer that is measured by SEM, and $\Delta C_W$ the water concentration difference between the salty water and fresh water. Here $\varepsilon = 0.021$ is the areal fraction of the freshwater layer (Supplementary Information section (4.7)) that is 1 nm thick[54], diffusing on the outer surface of the carbon fibers whose average diameter is 35 nm, with an average center-to-center separation of 65 nm. The latter values were obtained from the TEM pictures. By fitting the experimental data shown in Fig. 2(c), we obtained surface water diffusion constant of $D = 2.8\times10^{-5}$ cm$^2$/s at 20°C and $3.6\times10^{-5}$ cm$^2$/s at 80°C. These are in excellent agreement with the literature values ranging from $2.5\times10^{-5}$ cm$^2$/s to $6\times10^{-5}$ cm$^2$/s on a graphene surface over the similar temperature range[55].

**Salt rejection mechanism**

To understand the salt rejection mechanism, we used molecular dynamics simulations to show (see Fig. 3) that there is a freshwater layer between the salt water and the carbon surface, as well as between the salt water and its vapour. This is owing to (a) the formation of solvation shells in which each salt ion is enveloped by a layer of structured water molecules[56,57], thereby preventing the salt ions to be in direct contact with the solid surface, and (b) the layered structures formed by water molecules at the graphitic interface[58]. In the vicinity of the salt water's meniscus (Fig. 3(c)), water molecules that kinetically leave the freshwater layer can either be part of the vapour, or form islands/droplets on the carbon surface. The latter has been shown both experimentally and theoretically to be a favoured process even though the carbon surface is slightly hydrophobic[59]. This was attributed to the fact that most of the hydrogen bonds are satisfied in the island/droplet configuration, thereby lowering its overall energy. Such droplets/islands are then fast-transported on the carbon surface to form the freshwater flux, driven by the chemical potential gradient.

In Figure 3(a) we show the distribution of water and salt ions near the graphitic and vapour phase interfaces. Interestingly, on both sides there exists a thin layer that contains only water, with no salt ions. The layer near the graphene surface is about one atomic layer thick, while the layer near the air is about 2 to 3 atomic layers thick. Therefore in spite of the fact that the graphitic-salt water interface is slightly hydrophobic (the contact angle of water and salt water was measured to be about the same, with a value that is slightly larger than 90 degrees) and therefore non-wetting, there can be a pure water layer at the graphitic-salt water interface.

Apart from the water-vapour interfacial tension of ~58 mN/m, the pure water layer is also seen to form an interface with salty water, with a small surface tension of 4 mN/m (Fig. 3(b)) that prevents the mixing of the two. Based on this result, an interfacial sieving mechanism for high salt rejection is proposed as illustrated in Fig. 3(c). When salt water is in contact with the porous carbon membrane at a pressure below the LEP, a meniscus serving as a water-air interface is formed at the openings of the porous channels, in the region close to the interface with the porous ceramic substrate, where the pore size is the smallest and the areal carbon fiber density the highest. On the air side of the interface, water will form islands/clusters on the carbon fibers' outer surfaces[20,59]. These water clusters are supplied either from the surface layer of the bulk phase through surface diffusion or from condensation of vapour. In both cases the water clusters contain no salt. The formation of the water clusters is a kinetically equilibrium process, so the total number will be determined by the water concentration in the bulk phase as well as the temperature. The water clusters will then diffuse along the carbon surface, driven by the freshwater chemical potential gradient. Since the water diffusivity on carbon surface is high, surface diffusion dominates the total flux.

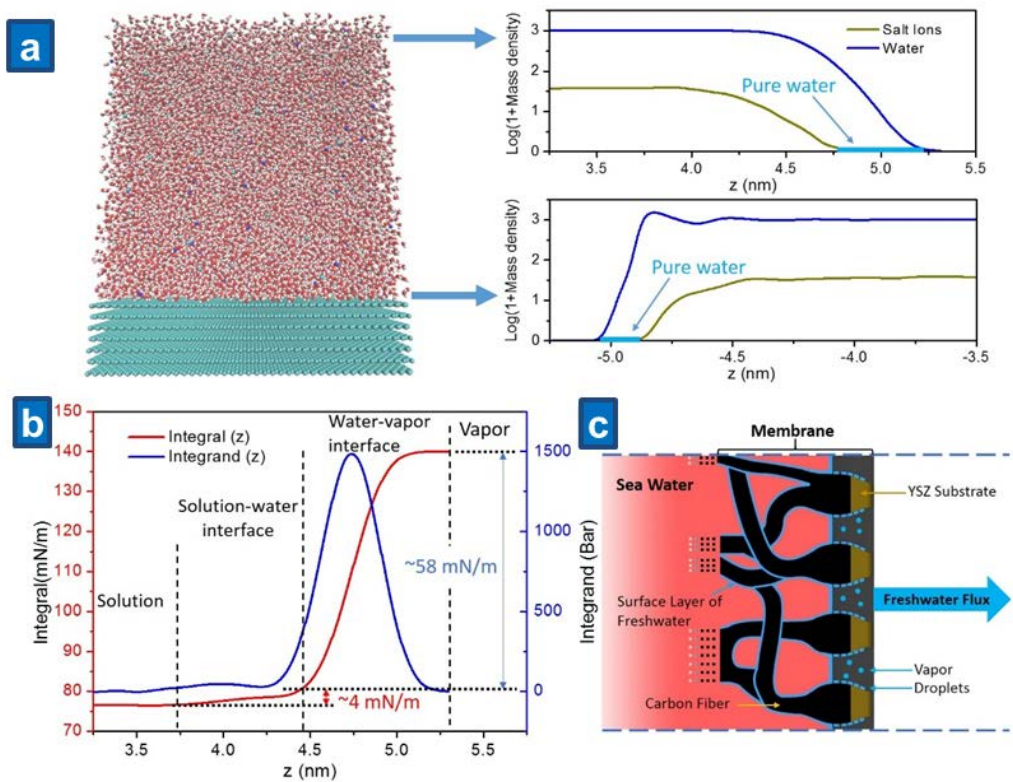

**Figure 3 | Desalination mechanism. (a)** Left panel: a molecular view of the simulated system. The concentration of the salt solution is about 3.5 wt%. Right panel: top figure shows the densities of salt ions (green line) and the water (blue line). Mass density is in units of g/liter. Two to three atomic layers of pure water is seen at the water-vapour interface, indicated by the light blue line. Bottom figure shows the same at the carbon-salt solution interface. A monolayer of pure water is seen to exist at the carbon surface. The surface water layer is noted to have lower density than the bulk. Center of the first carbon atomic layer is located at -5.35 nm. So there is a small "air gap" of around 3 Angstroms. **(b)** The blue line indicates the anisotropic component of the stress tensor whose integral (red line) gives the surface tensions. It is seen that besides the water-vacuum interfacial tension of 58 mN/m, there is a small interfacial tension between pure water and the saline solution that prevents mixing of the salt ions with the surface water layer. **(c)** An illustration of surface diffusion and the interfacial salt sieving effect. An explanation of the freshwater transport process is given in the text.

**Energy accounting**

Energy consumption counts significantly in the total desalination cost[60]. Membrane distillation involves vapour transport, which requires the expenditure of the latent heat energy. Since in the membrane distillation process using our carbon membranes more than 80% of water transported is attributed to surface diffusion, the energy consumption can be significantly reduced. To confirm this, the energy consumption was carefully studied in a setup schematically shown in Fig. 4. Temperatures denoted $T_1$, $T_2$, $T_3$ and $T_4$ were measured at different points as shown. The inlet temperature of the salt stream $T_3$ is close to the room temperature, while temperature $T_1$ is set to be higher than $T_3$, so that water will transport from the fresh water stream to the salty water stream by vapour diffusion as well as by surface diffusion. Thus the process used in our energy accounting measurements is similar to a FO process, but with an added temperature gradient. The parameters $h_0$ and $h_m$ shown in Fig. 4 are the heat transfer coefficients to account for the heat loss to the environment and the heat conduction between the two streams, respectively. The values of $h_0$ at different temperatures were determined by a separate experiment using an impermeable membrane (see Supplementary Information, section (5.1)). From the temperature data, measured in-flow and out-flow rates $P$, $F$, and membrane flux $V$ on the two sides of the membrane, plus the known values of latent heat and specific heats of water, an upper bound on the percentage $m$ of water transported through the membrane in vapor form, defined as $\bar{m} = m + (h_m \Delta T_m / VL)$, can be obtained (see Supplementary Information for details). Here $\Delta T_m$ denotes the average temperature difference across the membrane, $V$ is the transported freshwater flux across the membrane, and $L$ the latent heat. The results are listed in Table 1. They show that when the nanoporous carbon membrane was used, the upper bound $\bar{m} > m$ increases with $T_1$, but even at 80°C only ~10-20% water was transported as vapour and the rest was through surface diffusion. However, when a PTFE membrane was used, $\bar{m} > 100\%$ for temperatures over 30 °C. These results are consistent with the data shown in Fig. 2(b). It means that the intrinsic energy consumption of this process (FO plus temperature gradient) is reduced by at least 80% in the nanoporous carbon membrane as compared to the PTFE membrane. In Table 1 is also shown the order of magnitude difference in the transported freshwater fluxes between the two membranes. The increased temperature gradient is seen to have a larger effect on the PTFE membrane than the nanoporous carbon membrane, owing to the different (Knudsen diffusion dominant) transport mechanism.

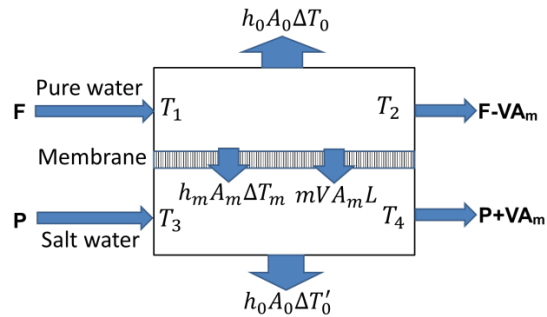

**Figure 4 | Energy accounting experiment.** A schematic illustration of the setup to measure the temperature change of the desalination process. The terms $h_0 A_0 \Delta T_0$ and $h_0 A_0 \Delta T_0'$ represent the heat loss to environment where $h_0$ and $A_0$ are the module heat transfer coefficient and surface area and $\Delta T_0$ and $\Delta T_0'$ the average temperature difference between the flow and the environment at the feed and permeate side, respectively. The term $h_m A_m \Delta T_m$ represents the heat conduction between the two streams where $h_m$ and $A_m$ are the membrane heat transfer coefficient and membrane area, respectively, and $\Delta T_m$ the average temperature difference between the two streams. *F* and *P* denote the constant flow rates of the pure water and salt water streams, respectively, and *V* is the transported freshwater flux from *F* to *P*. The term $mVA_m L$ represents the heat carried by vapor, where *m* is the percentage of vapour in the total membrane flux and *L* the latent heat. The energy accounting measurements can yield the value of *V* and an upper bound to the value of *m*.

Table 1: Upper bound on the fraction of transported water involving a phase change, and the total freshwater flux.

| $T_1$ (°C) | C-D35-2 membrane | | PTFE membrane | |
|---|---|---|---|---|
| | $\overline{m}$ | V(LMH) | $\overline{m}$ | V(LMH) |
| 30 | 3% | 43.4 | 64% | 0.98 |
| 40 | 6% | 46.5 | 208% | 1.24 |
| 50 | 9% | 51.1 | 249% | 1.81 |
| 60 | 12% | 56.7 | 313% | 2.57 |
| 70 | 15% | 62.8 | 231% | 4.64 |
| 80 | 18% | 69.5 | 186% | 7.69 |

## Conclusions

In summary, a nanoporous carbon composite membrane is found to display unprecedented high water flux in three membrane-based desalination processes. The fast transport of water is due to the extraordinary surface diffusion, while the excellent salt rejection rate is attributed to a unique interfacial sieving effect. This high-flux desalination mechanism, with no phase change, opens up the possibility of considerable energy savings for the desalination process, with the FO combined with a temperature gradient being a promising direction for its realization. The membrane fabrication process is simple, fast and easy to scale-up. Our findings can lead to the development of low cost, high-flux water desalination process(es), with the potential to alleviate the world-wide water crisis.

## Methods

**Experimental**

*Growth of carbon nanostructures on the hollow fiber*
YSZ hollow fibers were custom-made from YSZ nanoparticles (30 to 60 nm from Inframat Advanced Materials Co.) through a phase-inversion/sintering process[61,62]. The diameter of YSZ hollow fiber was about 0.91 mm with average pore size of 100 nm and porosity of 40%. The outer surface of YSZ hollow fiber was uniformly coated with nickel nanoparticles (20 to 30 nm, Fig. S1c) using a rotational sputtering deposition. A carbon layer was grown on nickel deposited YSZ hollow fiber through a catalytic chemical vapour deposition (CVD) process, in which acetylene was used as carbon source in the presence of hydrogen gas (acetylene to hydrogen volume ratio 1:10) to grown carbon nanowires at 700 $^{\circ}$C for 1 to 3 minutes. Then the CVD chamber was quickly cooled down to room temperature under argon flow. Following the same procedure a carbon composite membrane can also be grown on YSZ flat-sheet support.

*Membrane characterization*
Raman spectroscopy measurements were carried out on a Horiba Aramis confocal microprobe Raman instrument with He–Ne laser ($\lambda$ = 632.8 nm) at the outer surface of the C-ZrHF membranes. SEM images were taken by a FEI Nova Nano630 equipped with a focused ion beam (FIB), which facilitates in obtaining an ultra-smooth interface of C-DP-X membrane while preserving the initial structure. The elemental distributions of the membrane were analysed by energy dispersive X-ray (EDX) mapping in SEM. Transmission electron microscopy images were obtained by using a Titan ST microscope (FEI Co.), operating at 300 kV.

*Energy accounting experiment*

Carbon composite membranes grown on YSZ flat-sheet supports were used for energy accounting experiments in order to have large room to house the temperature probes on both sides of the membrane. Commercial porous PTFE membrane (W.L. Gore®) and dense polyethylene (PP) sheet were used as references for comparison. The membranes were mounted into a permeation cell made of polymethyl methacrylate (PMMA). Fresh water and draw solution (10 wt% NaCl) were recycled in each side of the membrane through circulation bathes. At each measurement point, the experiment was run for ~5 h to reach steady state, then the weight, conductivity, and temperatures at the inlet and outlet of each stream were recorded.

**Theory**

*Molecular dynamics simulations*

Molecular dynamics simulation was carried out by using the package GROMACS 4.6.7[63]. Parameterized force fields were adopted to describe the atomic interactions in the system[64-66]. The concentration of the NaCl solution was chosen to be ~3.5 wt%, similar to that of seawater. Carbon atoms were fixed at the crystallographic positions of the graphite lattice. All bonds of water molecules were constrained by using the SHAKE method[67]. The simulation was performed for 5 ns in the canonical ensemble with Berendsen thermostat[68] at a constant temperature of 300 K. The time step was set to be 1 fs. Long range electrostatic interactions were calculated with the particle mesh Ewald technique[69] and the van der Waals interactions were cut off at 1.2 nm. A custom GROMACS version based on GROMACS 4.5.5 was used to compute the 3D stress tensor from the simulated data[70].

6  Le Duc, Y. *et al.* Imidazole-quartet water and proton dipolar channels. *Angew. Chem. Int. Ed.* **50**, 11366–11372 (2011).

7  Kumar, M. *et al.* High-Density Reconstitution of Functional water channels into vesicular and planar block copolymer membranes. *J. Am. Chem. Soc.* **134**, 18631–18637 (2012).

8  Barboiu, M. & Gilles, A. From natural to bioassisted and biomimetic artificial water channel systems. *Acc. Chem. Res.* **46**, 2814–2823 (2013).

9  Skoulidas, A. I., Ackerman, D. M., Johnson, J. K. & Sholl, D. S. Rapid transport of gases in carbon nanotubes. *Phys. Rev. Lett.* **89**, 185901 (2002).

10 Hummer, G., Rasaiah, J. C. & Noworyta, J. P. Water conduction through the hydrophobic channel of a carbon nanotube. *Nature* **414**, 188–190 (2001).

11 Kalra, A., Garde, S. & Hummer, G. Osmotic water transport through carbon nanotube membranes. *Proc. Natl. Acad. Sci.* **100**, 10175–10180 (2003).

12 Hinds, B. J. et al. Aligned multiwalled carbon nanotube membranes. *Science* **303**, 62–65 (2004).

13 Majumder, M., Chopra, N., Andrews, R. & Hinds, B. J. Enhanced flow in carbon nanotubes. *Nature* **438**, 44–44 (2005).

14 Holt, J. K. et al. Fast mass transport through sub-2-nanometer carbon nanotubes. *Science* **312**, 1034–1037 (2006).

15 Whitby, M. & Quirke, N. Fluid flow in carbon nanotubes and nanopipes. *Nat. Nanotechnol.* **2**, 87–94 (2007).

16 Theresa, M., Pendergast, M. & Hoek, M. V. A review of water treatment membrane nanotechnologies. *Energy Environ. Sci.* **4**, 1946–1971 (2011).

17 Wu, J., Gerstandt, K., Zhang, H. B., Liu, J. & Hinds, B. J. Electrophoretically induced aqueous flow through single-walled carbon nanotube membranes. *Nat. Nanotechnol.* **7**, 133–139 (2012).

18 Verweij, H., Schillo, M. C. & Li, J. Fast mass transport through carbon nanotube membranes. *Small* **3**, 1996–2004 (2007).

19 Aluru, N. R. & Joseph, S. Why are carbon nanotubes fast transporters of water. *Nano Lett.* **8**, 452–458 (2008).

20 Muller, E. A. Purification of water through nanoporous carbon membranes: a molecular simulation viewpoint. *Curr. Opin. Chem. Eng.* **2**, 223–228 (2013).
17

energy, attained at 100% pump efficiency, is only 0.15% of the latent heat energy for water evaporation. Hence even by taking into account the inefficiency of the pump, it is still a small fraction of the total energy. Moreover, such peripheral energy cost is common to all membrane desalination processes.

**Acknowledgements**

Commercial PTFE membranes and FO membranes are provided by Dr. NorEddine Ghaffour and Dr. Tao Zhang from KAUST Water Desalination and Reuse Center. Z. L. wishes to acknowledge the support of KAUST URF/1/1723 grant and KACST RGC/3/1614 grant. P.S. wishes to acknowledge the support of KAUST Special Partnerships Award number UK-C0016, SA-C0040, HKUST grant SRFI 11/SC02, and William Mong Institute of Nanoscience and Technology grant G5537-E.

**Author contributions**

Z.L. and W.C. contributed to the initial ideas and experimental design. S.C. and P.S. contributed to the desalination mechanism and the relevant simulations and data analysis. W.C., Q. Z., and Z.F. carried out the experiments and S.C. carried out the simulation. K.W.H. contributed to data analysis. X.Z. contributed to characterization and data analysis. Z.L., W.C., S.C., P.S., and X.Z. participated in writing the manuscript.

**Competing financial interests**

The authors declare no competing financial interests.

# High-flux water desalination with interfacial salt sieving effect in nanoporous carbon composite membranes

Wei Chen[1], Shuyu Chen[2], Qiang Zhang[1], Zhongli Fan[1], Kuo-Wei Huang[1], Xixiang Zhang[1], Zhiping Lai[1*], Ping Sheng[2,3*]

[1]Division of Physical Science and Engineering, King Abdullah University of Science and Technology, Thuwal, Saudi Arabia

[2]Department of Physics, Hong Kong University of Science and Technology, Clear Water Bay, Kowloon, Hong Kong, China

[3]Institute for Advanced Study, Hong Kong University of Science and Technology, Clear Water Bay, Kowloon, Hong Kong, China

*Emails: zhiping.lai@kaust.edu.sa; sheng@ust.hk

**SUPPLEMENTARY INFORMATION**

**(1) Fabrication of carbon composite membrane on Yttrium-stabilized Zirconia (YSZ) hollow tube support**

Figure S1(a) shows the top-view SEM image of the outer surface of the YSZ hollow tube, and Fig. S1(b) the SEM image of its cross-section. The diameter of the YSZ hollow tube is 0.91 mm with an average pore size of 100 nm and porosity of 40%. The outer surface of the YSZ hollow tube was uniformly coated with nickel nanoparticles using rotational sputtering deposition. The diameter of the nickel nanoparticles is about 20 to 50 nm, visible in Fig. S1(c) as the small protrusions on the much larger zirconia particles. Subsequently the nickel coated YSZ hollow tube was placed inside a tubular furnace and heated to 700$^{\circ}$C in hydrogen with a hydrogen flow rate of 350 mL/min. When the temperature reached 700$^{\circ}$C, acetylene with a volume ratio of 1:10 to hydrogen gas was introduced to start the growth of carbon fibers. After 1 to 3 minutes, the furnace was opened fully to allow the temperature to quickly cool down to room temperature. The obtained sample was denoted as C-DP-X, where P represents the nickel deposition power in Watts and X denotes the growth time in minutes.

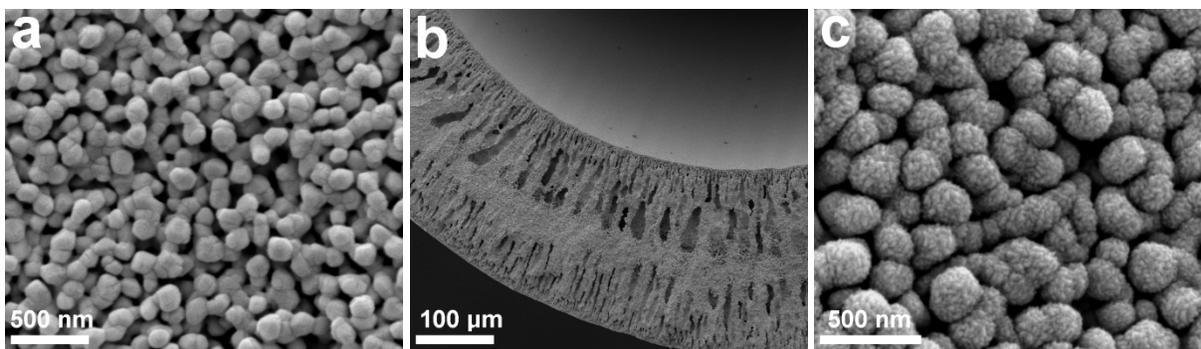

**Figure. S1| SEM images of the membrane.** **(a)** The outer surface of YSZ hollow tube, **(b)** cross-section of YSZ hollow tube wall, and **(c)** a top-view of YSZ hollow tube surface after nickel sputtering deposition. The nickel particles are visible as the small protrusions on the much larger zirconia particles.

**(2) EDX, TEM and Raman characterization of C-DP-X composite membranes**

The structure of the C-DP-X composite membrane was further studied by EDX, TEM and Raman spectroscopy. Raman spectroscopy measurements were carried out on a Horiba Aramis confocal microprobe Raman instrument with He–Ne laser (λ = 632.8 nm). SEM images were taken by a FEI Nova Nano630 equipped with a focused ion beam (FIB), which facilitates an ultra-smooth cut interface of the C-DP-X membrane while preserving its structure. The elemental distributions of membrane were analyzed by energy dispersive X-ray (EDX) mapping in SEM. TEM images were obtained using a Titan ST microscope (FEI Co.) operating at 300 kV.

Figure S2 shows the EDX elemental mapping of C, Ni and Zr at the interface between the carbon layer and the YSZ hollow tube support. The interface between carbon and YSZ support can be clearly identified because very few carbon and nickel atoms have penetrated into the support, while no Zr was found in the carbon layer. Nickel was deposited on the interface initially, but after growth it spreads over the entire carbon layer. Bright dots can be identified, indicating that nickel nanoparticles were lifted up by the growing carbon fibers. This observation agrees with the growth scenario proposed by Baker et al.[1] for carbon fibers grown on nickel surface.

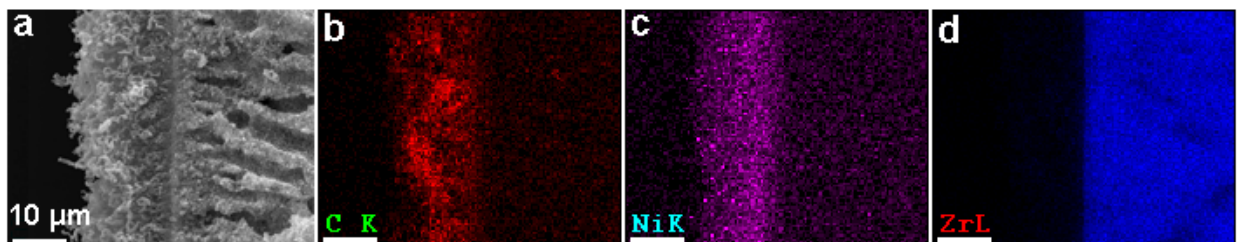

**Figure S2| SEM image and EDX mapping of the membrane.** **(a)** SEM image of C-D35-2, **(b)** EDX compositional mapping of carbon, **(c)** the same for Ni, and **(d)** the same for Zr. The scale bar is 10 μm.

The HRTEM image in Fig. 1(c) in the main text indicates that the fiber contains multilayers with an interlayer distance of 3.50 ± 0.05 Å, which is equal to the separation of graphitic layers. The inner channel possesses bamboo-knot-like structures with multi carbon layers (marked with arrow) that separate the channels into isolated compartments. A large number of carbon fibers were imaged at random, and the results showed that these knots existed in all the fibers without exception, indicating that the inner channels are not available for water transport. The graphitic structure of the carbon wall was further studied by Raman spectrum. As shown in Fig. S3, three bands centered at 1327, 1572 and 1605 cm$^{-1}$, corresponding to the characteristic D, G, and D' modes of graphitic carbon[2]. These Raman peaks have almost the same position, shape and relative ratio to those of carbon nanotubes, and hence confirm the multi-layered graphitic wall structure.

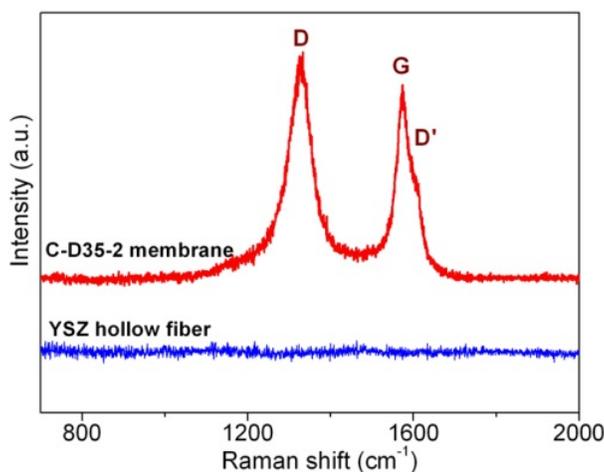

**Figure S3| Raman spectra of C-D35-2 composite membrane on YSZ hollow fiber support**.

**(3) Liquid entry pressure (LEP), gas permeation measurement, and pore size calculations**

LEP was measured based on the procedure recommended by Smolders and Franken[3]. The membrane was mounted on a glass module as illustrated in Fig. S4(a). Pure water was recycled through the inner channel of the C-D35-2 membrane and the conductivity was measured continuously by a conductivity electrode. 5 wt% salt water was fed into the outside chamber of the membrane from a pressurized tank. The pressure of the salt water was gradually increased by compressed nitrogen. The LEP value of the membrane was determined when a sharp increase was observed in the conductivity of the pure water stream. Figure S4(b) shows the LEP value to be inversely proportional to the membrane pore size. Therefore, the LEP of the membrane follows the Kelvin equation, $LEP = \frac{4\gamma}{d_P}|cos\theta|$, where $\gamma$ is the surface tension, $d_P$ the pore size and $\theta$ the contact angle.

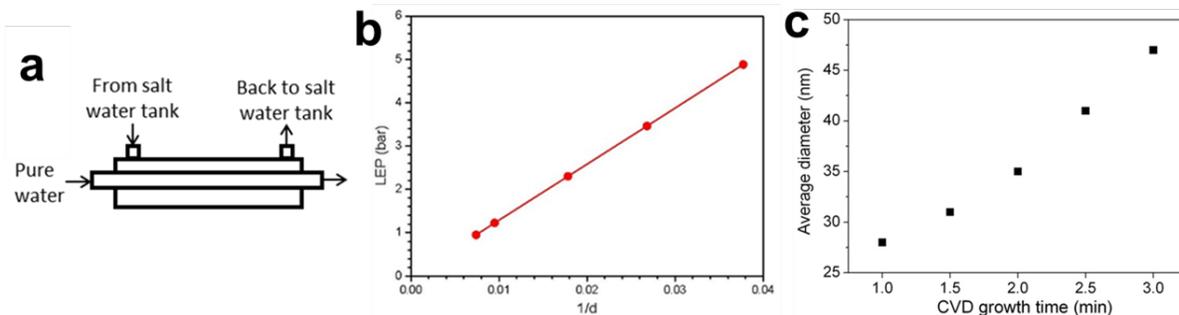

**Figure S4| Liquid entry pressure measurement. (a)** A schematic illustration of the setup for the LEP measurement. **(b)** LEP plotted as the reciprocal of the membrane pore size, determined from gas permeation. **(c)** The dependence of the average diameters of carbon fiber upon the CVD growth time.

Figure S5(a) shows the permeances of $H_2$, He, $CH_4$, $N_2$ and $CO_2$ through the C-D35-2 membrane at different trans-membrane pressure drops. The large permeance values indicate that the carbon composite membrane is highly permeable. Figure S5(b) shows that the gas permeance is inversely proportional to the square root of their molecular weight, implying that the gas transport mechanism through the composite membrane is dominated by Knudsen diffusion. The average pore size of the membrane can be estimated by the Yasuda-Tsai equation[4]:

$$d_P = \left(\frac{B_0}{K_0}\right)\left(\frac{16}{3}\right)\left(\frac{2RT}{\pi}\right)^{1/2} M^{-1/2},$$

where $d_P$ is the pore size, $B_o$ the geometric factor of the membrane which can be obtained from the slop of the permeation curves, $K_o$ the Knudsen permeability coefficient which can be obtained by extrapolation of the curves to zero pressure, R the gas constant, T the temperature, and M is the gas molecular weight. For example, after fitting the data, the values of $B_o$ and $K_0$ were found to be $8.57 \times 10^{-16}$ and $3.47 \times 10^{-6}$, respectively, for nitrogen permeating through the C-D35-2 membrane at 25°C. Hence the pore size is 31 nm in this case.

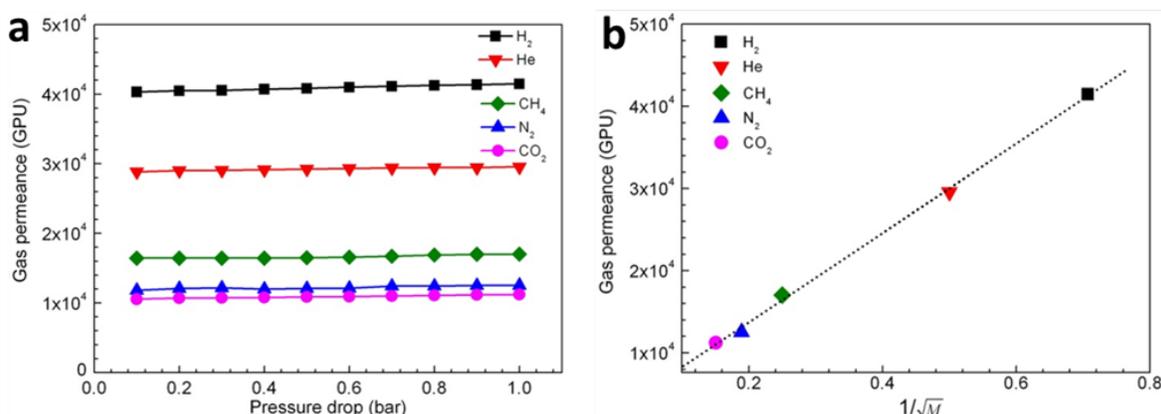

**Figure S5| Gas permeation measurements. (a)** Gas permeance measured at different trans-membrane pressure drop. **(b)** The permeance (at 1.0 bar) plotted as a function of the inverse square root of the gas molecular weight. The linear relationship confirms the Knudsen diffusion mechanism.

**(4) Desalination measurements for three different membrane processes**

## (4.1) Membrane distillation using synthetic seawater

An illustration of the vacuum membrane distillation setup in the contact model is shown in Fig. 2(a) in the main text, while a photo of the real setup is shown in Fig. S6(a). The same illustration is shown as Fig. S6(b). The C-D35-2 membrane was immersed into a salt solution with one end sealed by epoxy resin and the other end connected to a vacuum pump through a condensation trap. NaCl solutions were used as synthetic seawater. The solution was well-mixed by a stir bar, and the temperature was controlled by a heater. Vapor was drawn by vacuum and condensed first by a cooling water condenser followed by a liquid nitrogen trap (alternatively, cold water at 2° C was also used, with only 1% difference in collected water). The vacuum level is 825 Pa. The amount of the collected water was weighed at regular time intervals. The salt concentration of the collected water was measured by conductivity using a Thermo Scientific conductivity meter (equipped Orion® DuraProbe 013005MD conductivity electrode).

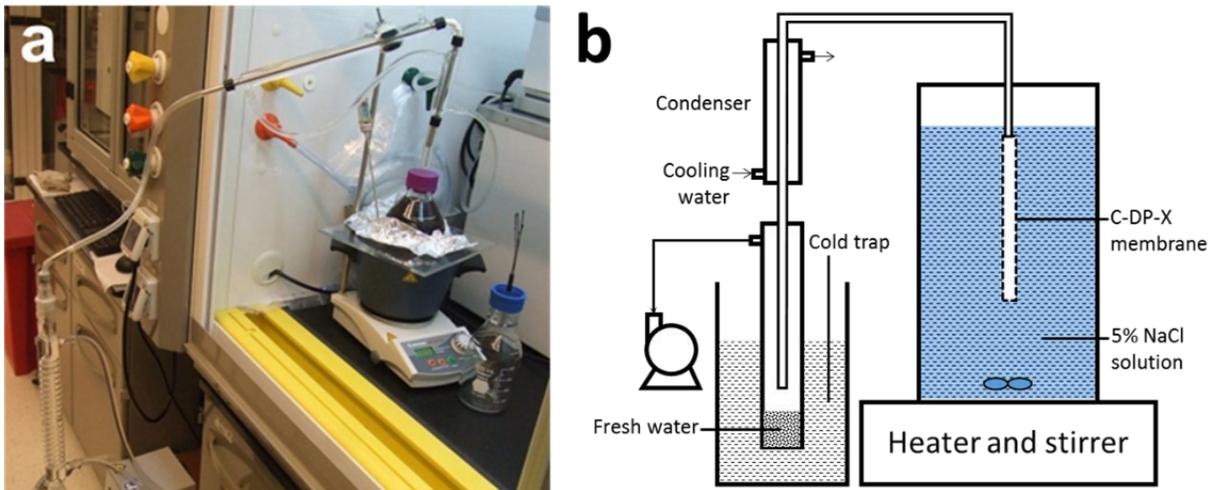

**Figure S6| The vacuum membrane distillation setup. (a)** A photograph of the vacuum membrane distillation setup in contact model for desalination of salt water. **(b)** Schematic illustration of the setup, same as Fig. 2(a) in the main text. The cold trap can use either liquid $N_2$ or simply cold water at 2°C.

In a typical VMD experiment, one C-D35-2 membrane with a length of 43.5 mm and diameter of 0.91 mm was subjected to VMD at 80 °C using 5 wt% NaCl solution as the synthetic seawater. After running for a duration of 6 h, 133.7 g of freshwater was collected. The fact that it was freshwater was determined by the resistance measurements. The membrane flux was calculated from the following equation,

$$J_w = \frac{m_w}{A\,t} = \frac{133.7 \times 10^{-3}}{3.14 \times 0.91 \times 10^{-3} \times 43.5 \times 10^{-3} \times 6} = 179.3 \text{ LMH} ,  \tag{S1}$$

where $J_w$ denotes the water flux, $m_w$ the amount of collected freshwater, A the effective membrane area, t the collection duration. From this simple calculation, a freshwater flux of 179.3 LMH is obtained.

## (4.2) Membrane distillation of seawater from the Red Sea

Red Sea seawater was taken from the KAUST south beach. Table S1 lists the composition of the major ions. The total salinity is about 4.1wt%, higher than the average seawater salinity. By using real seawater, we first conducted the VMD experiment over a single C-D35-2 hollow tube membrane. The performance is shown by dark square symbols in Fig. S7. Once again, more than 99.9% salt rejection was achieved. The freshwater flux is very close to that when 5 wt% NaCl was used as synthetic seawater, i.e., somewhat lower if compared to the flux when 4.1wt% of synthetic seawater is used. The reason is due to the significant amount of divalent ions present in the real seawater, which have a higher osmotic pressure so that the equivalent (synthetic seawater) salinity is higher. It is to be noted that the same reduction in the freshwater flux was observed for commercial membranes in the presence of divalent ions[5,6]. However, in actual seawater desalination processes most of these divalent ions are removed by pre-treatments.

Hollow YSZ tube module has the advantage of simplicity in scaling-up because it just involves increasing the number of YSZ hollow tubes in one module. To demonstrate this capability, we have bound three C-D35-2 membranes in one module. The water flux (which is normalized by the total membrane area) over this triple bundle is very close to that over single membrane, as shown by the green square symbols in Fig. S7. Hence the total amount of freshwater is essentially tripled over the single YSZ tube membrane, for the same time duration. This result indicates the good reproducibility of our membrane fabrication process.

**Table S1| Major ion composition of Red Sea water at Jeddah[7].**

| Ions | Concentration (g/L) |
|---|---|
| Chloride ($Cl^-$) | 22.219 |
| Sodium ($Na^+$) | 14.255 |
| Sulfate ($SO_4^{2-}$) | 3.078 |
| Magnesium ($Mg^{2+}$) | 0.742 |
| Calcium ($Ca^{2+}$) | 0.225 |
| Potassium ($K^+$) | 0.21 |
| Bicarbonate($HCO_3^-$) | 0.146 |
| Bromide ($Br^-$) | 0.072 |
| **Total dissolved solids (TDS)** | 40.947 |

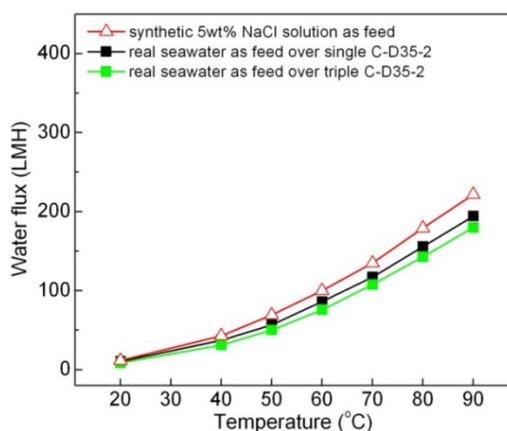

**Figure S7| Freshwater flux of C-D35-2 membrane using the Red Sea seawater.** The red open triangles denote the water flux on 5 wt% NaCl synthetic seawater. The dark and green solid squares denote the water flux for the Red Sea seawater taken from KAUST south beach. The triple C-D35-2 membrane means three C-D35-2 membranes bound in one module. The total amount of freshwater collected over the triple C-D35-2 was about three times of that over single C-D35-2, so the membrane flux remains almost the same.

### (4.3) Effect of divalent ions

Figure S8 shows the preliminary results of the membrane fluxes in $Na_2SO_4$, $CuCl_2$ and $MgSO_4$ salt solutions. The weight percentage of all salts were fixed at 1 wt%. The membrane flux again shows a surface diffusion behavior in all of these salts, but it decreases significantly for divalent ions as compared to NaCl. This can be clearly seen in Fig. S8(b) where the flux is normalized by the osmotic pressure of the salt solution. This result is consistent with what was observed previously with commercial membranes[5,6]. However, the concentration of divalent ions in real seawater is low, so overall the decline in the membrane flux is minor, as proved by our experimental results using the Red Sea seawater. Furthermore, in actual desalination processes, most of these divalent ions will be removed during pretreatments.

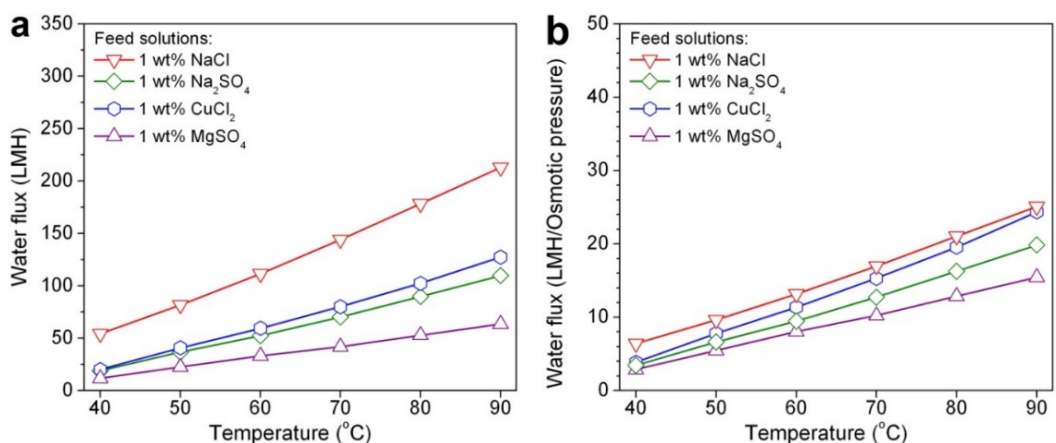

**Figure S8| Water flux of the C-D100-2 membrane in different salts. (a)** Membrane flux at different temperatures in 1 wt% of $CuCl_2$, $Na_2SO_4$ and $MgSO_4$ aqueous solutions. The data of 1 wt% NaCl solution in Fig. 2B (main text) is used for comparison. **(b)** The membrane flux normalized by the osmotic pressure

of the salt solutions.

**(4.4) Reverse osmosis**

Due to the limit of the liquid entry pressure, the RO process over the carbon composite membrane is limited to brackish water desalination. The setup for reverse osmosis is similar to that used for the LEP measurements. The C-D35-2 membrane was mounted on a glass module. Pure water was recycled through the inner channel of the C-D35-2 membrane and the conductivity was measured by a conductivity electrode. Salt water was recycled through the chamber outside the membrane from a pressurized tank. A pressure of 3.0 bar was applied to the salt water stream to push the water across the membrane to the pure water stream. The temperature of both streams was maintained to be the same, so as to avoid vapour transport through the membrane. The concentration of the salt water was 2000 ppm, i.e., in the brackish regime. The conductivity of the pure water stream was below 1.1 µS/cm (equivalent to 1 ppm salt concentration) even after two-days running in all of our RO measurements, indicating over 99.9% salt rejection rate. Figure S9 shows the specific water flux of the C-D35-2 membrane increases from 11.9 to 28.5 LMH/bar when the operation temperature was increased from 20 to 80°C.

The specific water flux shown in Fig. S9 was calculated as follows. In a typical RO experiment, one C-D35-2 membrane with a length of 52.3 mm and diameter of 0.91 mm was subjected to RO at 80 °C under 3.0 bar, and 2000 ppm NaCl solution was used as the feed stream. After running for a duration of 12 h, the amount of water in the freshwater stream increased by 69.36 g. From Eq. (S1), the freshwater flux is calculated to be 38.7 LMH. The specific water flux is then evaluated by the following formula:

Specific flux = flux/(applied pressure–osmotic pressure) = $\dfrac{38.7}{\left(3.0-\dfrac{2000\times2\times8.314\times293.15}{58.5\times101325}\right)}$ = 28.5 LMH/bar

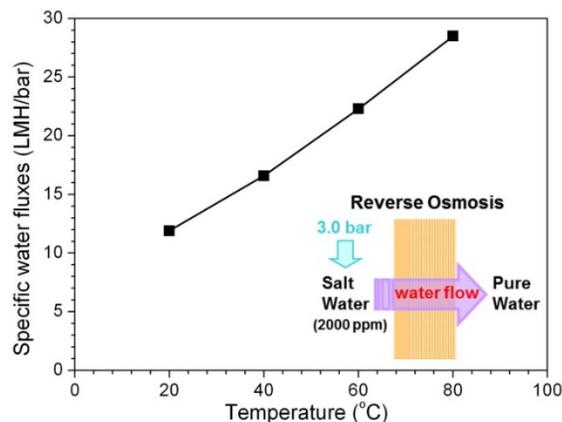

**Figure S9| Water flux of the reverse osmosis process.** The specific water flux at different temperatures. The temperatures on both sides of the membrane were maintained the same in all tests.

## (4.5) Forward osmosis

The setup for forward osmosis is similar to RO, except no pressure is applied to the salt water stream. So in this case freshwater transports from the pure water stream to the salt water stream driven by the osmotic pressure difference. The salt water is called draw solution. Figure S10(a) shows the FO water flux on a commercial PTFE membrane at different salinity of the draw solution. Compared to the water flux of the carbon composite membrane that is shown in Fig. 2C in the main text, the water flux of the PTFE membrane is much smaller. Figure S10(b) shows the water flux of the carbon composite membrane when the pure water stream was replaced by 3.5 wt% NaCl solution, while the salinity of the draw solution is changed from 5 to 10 wt%. The results showed that the carbon composite membrane is able to extract freshwater from salt water with a very high flux.

An example to calculate the water flux in FO is illustrated below. In a typical FO experiment, one C-D35-2 membrane with a length of 43.8 mm and diameter of 0.91 mm was subjected to FO at 80 °C. About 1 liter 10 wt% NaCl solution was used as the draw solution to extract water from a pure water stream. After running for a period of 24 h, the amount of the pure water stream decreased by 221.18 g. The same amount of weight increase was found in the draw solution stream. From Eq. (S1), the water flux is therefore calculated to be 73.6 LMH. That is the result shown in Fig. 2(c) in the main text. In Fig. S10(b) below, we have replace the feed by a 3.5wt% NaCl solution. The freshwater flux is decreased to around 18 LMH, but still much higher than that shown in Fig. S10(a) for the PTFE membrane with pure water as the feed.

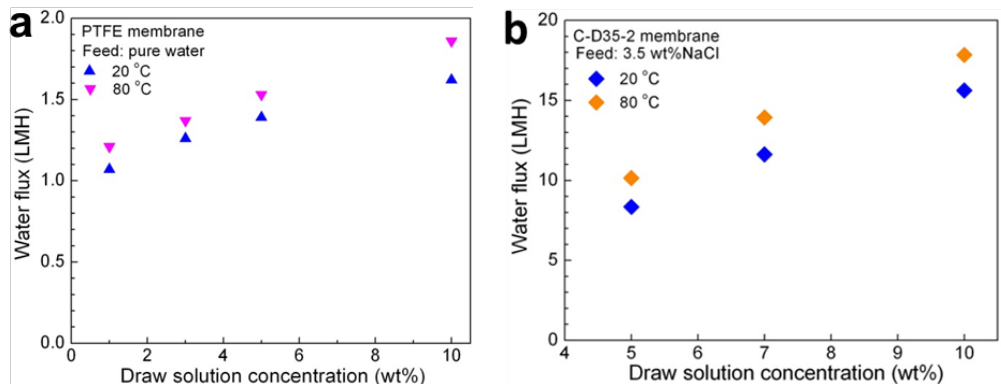

**Figure S10| Water flux comparison in the forward osmosis process. (a)** Water flux of PTFE membrane at different temperatures and different salinity of the draw solution in the FO process. **(b)** Water flux of the C-D35-2 membrane when the pure water stream was replaced by 3.5 wt% NaCl solution. The temperatures on both sides of the membrane were maintained to be the same in all the measurements. It is obvious that the water flux in (b) is much larger than that in (a). However, because the feed in (b) is 3.5 wt% salt solution, the freshwater flux is less than that shown in Fig. 2(c) in the main text.

## (4.6) Stability of the C-DP-X membrane

Yttrium-stabilized zirconia (YSZ) is a very stable ceramic, while carbon is also chemically inert, so the C-DP-X composition is very stable. To confirm the stability of its functionality, one C-D100-2 membrane was tested under the FO process at 80 °C for an extend period of 168 h. As shown in Fig. S11(a), the water flux kept almost constant during this extended period. For another C-D100-2 membrane that was tested by VMD studies 8 months ago and stored under ambient conditions thereafter, the membrane was tested by VMD again. Fig. S11(b) shows that the two test results are almost identical, which indicates the membrane property is very stable.

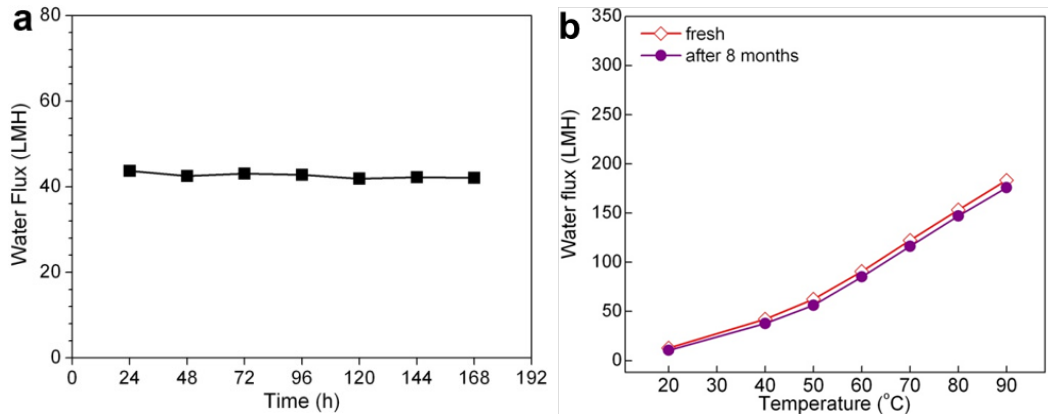

**Figure S11| Membrane stability studies. (a)** A C-D100-2 membrane was tested in the FO mode at 80 °C continuously for 168 h. 3.5 wt% NaCl solution was used as draw solution to extract water from pure water stream, and both draw solution and feed were renewed every 24 h. **(b)** A C-D100-2 membrane was tested by VMD in 3.5 wt% NaCl and the test was repeated after 8 months.

### (4.7) Estimation of the surface diffusion constant

Since the surface diffusion flux of freshwater is transported on the outer surface of the carbon fibers, we first estimate the areal density of the fibers, based on the geometric relationship illustrated in Fig. S12. From the TEM image Fig. 1(c) in the main text, the average diameter of carbon fibers is around 35 nm. The average gap between the fibers is 30 nm. Based on a dense packed model, the areal density of the fibers is around 12%. For a 1 nm layer of freshwater on the outer surface of the carbon fibers, the net areal fraction of the freshwater layer would be 12% times 0.1719, or about 0.021.

It is to be noted that in this FO process, the salt concentration gradient should be able to drive the salt ions to diffuse across to the pure water side. However, no salt concentration was detected in all the FO experiments. It shows that our membrane is impermeable to the salt ions.

Diffusivity was calculated as follows. The slope of curves in Fig. 2(c) is equal to $J_W/\Delta C_W$. The unit of $J_w$ is LMH, the unit of $\Delta C$ in Fig. 2(c) is 10 kg/m$^3$. So the conversion factor to m/s is 1/36000. The fitted slope for flux at 80°C in Fig. 2(c) is 5.5, which is equivalent to $1.53\times10^{-4}$ m/s. From equation (1) in the text, we have

$$D = \frac{J_W}{\Delta C_W}\frac{\delta}{\varepsilon} = 3.64\times 10^{-9}\, m^2/s = 3.64\times 10^{-5}\, cm^2/s$$

The conclusion of this subsection is that the observed high freshwater flux in our nanoporous carbon composite membrane is consistent with the previous literature values of the diffusion constant of water on carbon surfaces, and therefore an intrinsic property of the carbon membrane.

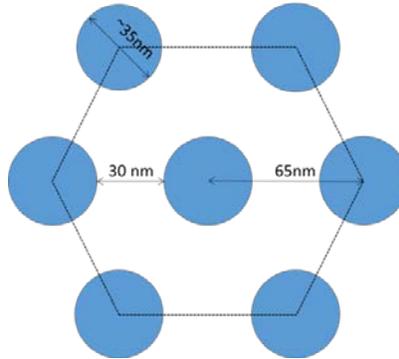

**Figure S12| A schematic illustration of the geometric relationship for the cross section of the dense carbon fibers in the C-D35-2 membrane.** Blue circles denote the carbon fibers. Surface diffusion of the freshwater flux is along the surfaces of the carbon fibers.

### (5) Energy accounting measurements

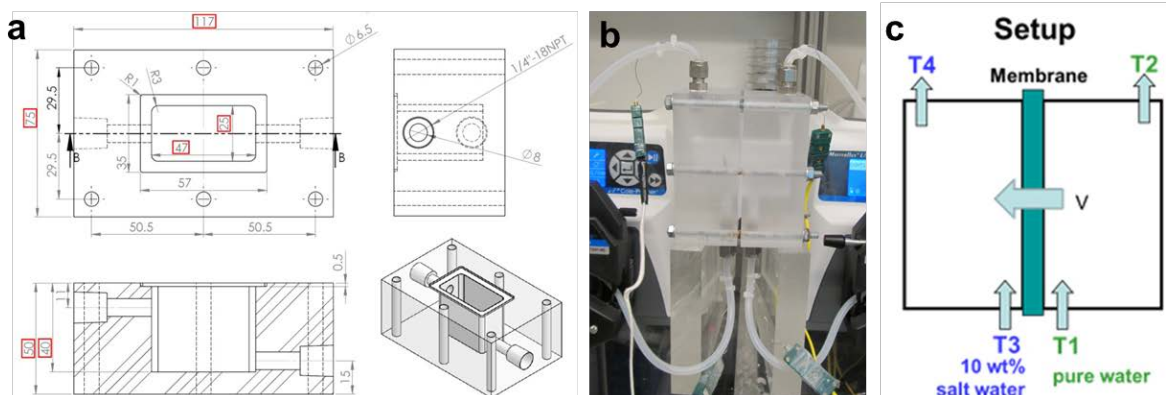

**Figure S13| Details of the energy accounting measurement setup. (a)** The drawings of the PMMA modules. The right and left modules are the same, hence only half of the setup is shown. **(b)** A photo of the setup. **(c)** The schematics of the setup and the relevant input parameters for the energy balance calculation.

The following parameters are involved in the energy accounting measurements.

*F*:        Flow rate of pure water stream, which is fixed to be 0.939 g/s in all energy counting experiments.

*P*:        Flow rate of salt water stream, which is fixed to be 0.987 g/s in all energy counting experiments.

$C_p$: Heat capacity of pure water (4.18 J/g).
$C_P'$: Heat capacity of 10 wt% salt water (3.67 J/g).
$V$: Total amount of water transport across the membranes, g/s.
$m$: Percentage of water that transport across the membranes in vapor form.
$L$: Latent heat of water, J/g.
$A_m$: membrane area (1.175x10$^{-3}$ m$^2$).
$h_m$: Heat transfer coefficient to count the heat exchange between the two streams through the membrane, J/(m$^2$·K·s).
$\Delta T_m$: Average temperature difference between the two streams, °C.
$A_o$: Inner surface area of the module (6.935x10$^{-3}$ m$^2$).
$h_o$: Heat transfer coefficient of the module to count heat loss to the environment, J/(m$^2$·K·s).
$\Delta T_o$: Temperature difference between pure water stream and environment, °C;
$\Delta T_o'$: Temperature difference between salt water steam and the environment, °C.
$T_{ref}$: Reference temperature, here taken to be $T_{ref}$ = 0 °C.
$\overline{H}_W^0$: Standard enthalpy of pure water at the reference temperature, taken to be 0 °C.
$\overline{H}_S^0$: Standard enthalpy of the salt water stream at the reference temperature, taken to be 0 °C.
$T_{AV}$: Average temperature at the fresh water side $(T_1+T_2)/2$.
$T_{AV}'$: Average temperature at the salt water side $(T_3+T_4)/2$.

Energy balance at the pure water side is given by

$$H_1 = H_2 + H_m + H_0 + H_V \ , \tag{S2}$$

where $H_1$ is the enthalpy of the feed, $H_2$ is the enthalpy of exit, $H_m$ is the heat exchange by membrane conduction, $H_0$ is the heat loss to the environment, and $H_v$ is the energy carried by the transported water. We can write each term in Eq. (S2) as follows:

$$H_1 = F\overline{H}_W^0 + FC_P(T_1 - T_{ref}) \ , \tag{S3}$$

$$H_2 = (F - VA_m)\overline{H}_W^0 + (F - VA_m)C_P(T_2 - T_{ref}) \ , \tag{S4}$$

$$H_m = h_m A_m \Delta T_m \ , \tag{S5}$$

$$H_0 = h_0 A_0 \Delta T_0 \ , \tag{S6}$$

$$H_V = VA_m \overline{H}_W^0 + mVA_m L + VA_m C_P(T_{AV} - T_{ref}) \ . \tag{S7}$$

The overall energy balance on the freshwater side can therefore be arranged as

$$m + \frac{h_m \Delta T_m}{VL} = \frac{FC_P(T_1-T_2) - VA_m C_P(T_{AV}-T_2) - h_0 A_0 \Delta T_0}{VA_m L} \quad . \tag{S8}$$

In Eq. (S8), every term on the right hand side is either measured in the experiment or previously known (such as the latent heat of water *L* that is taken at $T_{AV}$). On the left hand side the quantity *m* is what we would like to know, since it represents the fraction of water that is transported through the membrane in vapour form.

The energy balance on the salt water side can be similarly written as

$$H_3 + H_m + H_V = H_4 + H_0' \quad , \tag{S9}$$

where

$$H_3 = P\overline{H}_S^0 + PC_P'(T_3 - T_{ref}) \quad , \tag{S10}$$

$$H_m = h_m A_m \Delta T_m \quad , \tag{S11}$$

$$H_V = VA_m \overline{H}_W^0 + mVA_m L + VA_m C_P(T_{AV}' - T_{ref}) \quad , \tag{S12}$$

$$H_0' = h_0 A_0 \Delta T_0' \quad , \tag{S13}$$

$$H_4 = P\overline{H}_S^0 + PC_P'(T_4 - T_{ref}) + VA_m \overline{H}_W^0 + VA_m C_P(T_4 - T_{ref}) \quad . \tag{S14}$$

Equation (S14) is obtained by assuming the salt water to be close to an ideal solution, so that the enthalpy can be calculated from the enthalpy of the two streams, i.e., the feed salt water stream *P* and the water flux *V* transported through the membrane. The overall energy balance on the salt water side can therefore be arranged as,

$$m + \frac{h_m \Delta T_m}{VL} = \frac{PC_P'(T_4-T_3) + h_0 A_0 \Delta T_0' - VA_m C_P(T_{AV}'-T_4)}{VA_m L} \quad . \tag{S15}$$

Again, the right hand side quantities in Equation (S15) are either previously known or measured in the experiment. The *m* on the left hand side is what we are after. Since $h_m$ is not accurately known, we define $\overline{m} = m + \frac{h_m \Delta T_m}{VL}$ to be the upper bound of *m*, as $\frac{h_m \Delta T_m}{VL}$ is always positive. Equations (S8) and (S15) represent two equations for determining the upper bound of *m* from the pure water side and from the salt water side. To minimize the experimental error, we choose to evaluate the average of $\overline{m}$ by adding Eqs. (S8) and (S15) and dividing by two:

$$\overline{m} = m + \frac{h_m \Delta T_m}{VL} = \frac{FC_P(T_1-T_2) + PC_P'(T_4-T_3) - VA_m C_P(T_{AV}+T_{AV}'-T_2-T_4) - h_0 A_0 \Delta T_0 + h_0 A_0 \Delta T_0'}{2VA_m L} \quad . \tag{S16}$$

Below we first use a dense polypropylene (PP) membrane, which is impermeable to water, to separately determine the heat loss to the environment on the two sides of the membrane. Then we carry out the experimental energy accounting measurements on a porous PTFE membrane and a carbon composite membrane grown on flat sheet YSZ support (denoted as C-Zr-sheet-2). The thickness of the PP membrane, the C-Zr-sheet-2 membrane, and the PTFE membrane are 0.60 mm, 0.66 mm and 0.18 mm, respectively. Pure water and draw solution (10 wt% NaCl) were recycled at each side of the membrane through circulation bathes. At each measurement point, the experiment was run for ~5 h to reach steady state, then the weight, conductivity, and temperatures at the inlet and outlet of each stream were recorded.

### (5.1) Control experiment using impermeable PP sheet

Impermeable PP membrane was first used to obtain the heat transfer coefficient of the setup, $h_o$, by using the following equation that is simplified from Eqs. (S2) and (S9) with $V=0$ and $m=0$,

$$F(T_1 - T_2)C_P = h_m A_m \Delta T_m + h_o A_o \Delta T_o ,  \tag{S17}$$

$$P(T_3 - T_4)C_P' + h_m A_m \Delta T_m = h_o A_o \Delta T_o' .  \tag{S18}$$

By adding the two equations, we obtain the following relation:

$$h_o = \frac{F(T_1-T_2)C_P + P(T_3-T_4)C_P'}{A_o(\Delta T_o + \Delta T_o')} .  \tag{S19}$$

The recorded temperatures are listed in Table S2. The heat transfer coefficient, $h_o$, which counts the heat loss to the environment, can thus be obtained from equation (S19). At similar inlet and outlet temperatures the $h_o$ values should be the same when the PP membrane is replaced by either PTFE or C-Zr-sheet-2 membranes. Since the lab environment is almost static, so $h_0 \approx \lambda_{PMMA}/d_0$, where $\lambda_{PMMA}$ is the thermal conductivity of the module that is made from PMMA, and $d_0$ the wall thickness. The calculated $\lambda_{PMMA}$ is in the range of 0.12 ~ 0.38 W/(m·°C), which is comparable to the reported PMMA values (0.16 ~ 0.30 W/(m·°C))[8-10].

**Table S2| Temperatures recorded in energy accounting of PP sheet.**

| T1 (°C) | T2 (°C) | T3 (°C) | T4 (°C) | $\Delta T_m$ (°C) | $\Delta T_o$ (°C) | $\Delta T_o'$ (°C) | $h_o$ (J/(m²·K·s)) |
|---|---|---|---|---|---|---|---|
| 30.0 | 29.8 | 25.3 | 25.4 | 4.55 | 7.70 | 3.15 | 5.6±0.1 |
| 40.0 | 39.4 | 26.0 | 26.4 | 13.50 | 17.50 | 4.00 | 6.1±0.1 |
| 50.0 | 48.8 | 26.1 | 26.7 | 23.00 | 27.20 | 4.20 | 11.7±0.2 |
| 60.0 | 58.1 | 26.2 | 27.1 | 32.40 | 36.85 | 4.45 | 14.7±0.2 |
| 70.0 | 67.5 | 26.3 | 27.4 | 41.90 | 46.55 | 4.65 | 16.4±0.2 |
| 80.0 | 76.8 | 26.5 | 27.9 | 51.20 | 56.20 | 5.00 | 17.7±0.3 |

### (5.2) Energy accounting using the C-Zr-Sheet-2 membrane and the PTFE membrane

The latent heat (*L*) of seawater does not change much with salinity, but has the following relationship with temperature[11],

$$L = 2500.39 - 2.3683T + 4.31 \times 10^{-4}T^2 - 1.131 \times 10^{-5}T^3 \text{ J/g}$$

Temperatures recorded on porous PTFE and C-Zr-sheet-2 membranes are listed in Tables S3 and S4, respectively. With the $h_0$ values obtained from PP membranes, the upper bound percentage, $\bar{m}$, of water that transport through membrane in vapor form can be obtained from Eq. (S16). These values are shown in the right columns of Tables S3 and S4. For the PTFE membrane, $\bar{m}$ is larger than 100% except at the fresh water inlet temperature of 30 °C, since at this condition the temperature difference is very small. Here $\bar{m} > 100\%$ is allowed since it represents an upper bound to $m$. The values of $\bar{m}$ shown in Table S3 are completely consistent with the data shown in the inset of Fig. 2(b), where the MD using PTFE membrane show very good agreement with the predicted Knudsen diffusion water flux. For the carbon composite membrane, however, $\bar{m}$ is seen to be less than 20% even at 80 °C. Since $m < \bar{m}$, it is therefore confirmed that more than 80% of water is transported by surface diffusion, which is consistent with the MD and FO results shown in Figs. 2(b) and 2(c).

**Table S3| The data obtained in energy accounting of the PTFE membrane**.

| T1 (°C) | T2 (°C) | T3 (°C) | T4 (°C) | ΔT$_m$ (°C) | ΔT$_o$ (°C) | ΔT$_o$' (°C) | V (LMH) | h$_o$ (J/(m²·K·s)) | $\bar{m}$ % |
|---|---|---|---|---|---|---|---|---|---|
| 30.0 | 29.8 | 25.8 | 25.9 | 4.05 | 7.70 | 3.65 | 0.98±0.01 | 5.6±0.1 | 64% |
| 40.0 | 39.1 | 25.9 | 26.2 | 13.50 | 17.35 | 3.85 | 1.24±0.02 | 6.1±0.1 | 208% |
| 50.0 | 48.3 | 26.1 | 26.7 | 22.75 | 26.95 | 4.20 | 1.81±0.03 | 11.7±0.2 | 249% |
| 60.0 | 57.4 | 26.2 | 27.7 | 31.75 | 36.50 | 4.75 | 2.57±0.06 | 14.7±0.2 | 313% |
| 70.0 | 66.4 | 26.4 | 28.3 | 40.85 | 46.00 | 5.15 | 4.64±0.12 | 16.4±0.2 | 231% |
| 80.0 | 75.5 | 26.3 | 29.1 | 50.05 | 55.55 | 5.50 | 7.69±0.22 | 17.7±0.3 | 186% |

**Table S4 | The data obtained in energy accounting of the C-Zr-Sheet-2 membrane**.

| T1 (°C) | T2 (°C) | T3 (°C) | T4 (°C) | ΔT$_m$ (°C) | ΔT$_o$ (°C) | ΔT$_o$' (°C) | V (LMH) | h$_o$ (J/(m²·K·s)) | $\bar{m}$ (%) |
|---|---|---|---|---|---|---|---|---|---|
| 30.0 | 29.7 | 25.5 | 25.7 | 4.25 | 7.65 | 3.40 | 43.40±2.15 | 5.6±0.1 | 3% |
| 40.0 | 39.1 | 25.7 | 26.1 | 13.65 | 17.35 | 3.70 | 46.47±2.46 | 6.1±0.1 | 6% |
| 50.0 | 48.4 | 25.9 | 26.6 | 22.95 | 27.00 | 4.05 | 51.06±2.86 | 11.7±0.2 | 9% |
| 60.0 | 57.6 | 26.2 | 27.3 | 32.05 | 36.60 | 4.55 | 56.68±3.34 | 14.7±0.2 | 12% |
| 70.0 | 66.7 | 26.5 | 28.2 | 41.00 | 46.15 | 5.15 | 62.81±3.87 | 16.4±0.2 | 15% |
| 80.0 | 75.8 | 26.8 | 29.1 | 49.95 | 55.70 | 5.75 | 69.45±4.43 | 17.7±0.3 | 18% |

We would also like to note that there is a very large difference in the volume *V* of the freshwater being transported across the membrane in the two experiments, with the C-Zr-Sheet-2 membrane showing a value of *V* that is a factor of 9 to 40 times that for the PTFE membrane (see Tables S3 and S4). In particular, *V* for the C-Zr-Sheet-2 membrane is noted to range from 1.5% to 2.4% of *F* and *P*, the constant inlet feed fluxes of the two streams. This percentage is well within the accuracy of the present experiment, with the weights of the amount of water determined to within one part in $10^4$. Such a high membrane flux is due to the relatively open structure of the C-Zr-Sheet-2 membrane, as compared to the PTFE membrane, which is dense. The conclusion that *m*<18% also cannot be due to errors in our temperature measurements. In fact, by assuming *m*=100% in our membrane, the measured temperatures $T_2$ and $T_4$ ($T_1$ and $T_3$ are controlled by the circulation chiller) would deviate by ~10° C from the measured values. Since the accuracy of our temperature measurements is 0.1° C, such errors are impossible. Thus the energy accounting experiments confirm not only the existence of a significant fraction of the freshwater flux being due to surface transport in the C-Zr-Sheet-2 membrane without involving a phase change, but also the very large net freshwater flux exhibited by our nanoporous carbon composite membranes.